\documentclass[final,5p,times,twocolumn]{elsarticle}
\usepackage[utf8]{inputenc}
\usepackage[T1]{fontenc}
\usepackage{amsmath}
\usepackage{url}
\usepackage{array}
\usepackage{bbm}
\usepackage{xspace}
\usepackage{siunitx}
\biboptions{sort&compress}
\usepackage[hidelinks]{hyperref}

\usepackage{algorithm,algpseudocode}
\usepackage{xcolor}
\usepackage{listings}
\usepackage{tcolorbox}
\usepackage{nicefrac}
\usepackage{placeins}

\usepackage{mathtools}
\usepackage{lipsum}
\usepackage{braket}
\newcommand{\tabitem}{~~\llap{\textbullet}~~}

\lstdefinestyle{mystyle}{breakatwhitespace=false,         
breaklines=true,}

\def\ContinueLineNumber{\lstset{firstnumber=last}}

\usepackage{lineno}
\modulolinenumbers[5]

\usepackage{caption}
\newcommand{\simos}{SimOS\xspace}

\makeatletter
\def\ps@pprintTitle{%
	\let\@oddhead\@empty
	\let\@evenhead\@empty
	\let\@oddfoot\@empty
	\let\@evenfoot\@oddfoot
}
\makeatother

\begin{document}

\begin{frontmatter}
\renewcommand\thesubsection{\Alph{subsection}}
\title{\simos: A Python Framework for Simulations of  Optically Addressable Spins}

\author{Laura A. V\"olker\fnref{ETH}}
\author{John M. Abendroth\fnref{ETH}}
\author{Christian L. Degen\fnref{ETH}}

\author{Konstantin Herb \fnref{ETH}\corref{correspondingAuthor}}
\address{Department of Physics, ETH Z\"urich, Otto-Stern-Weg 1, 8093 Z\"urich, Switzerland}

\cortext[correspondingAuthor]{Corresponding author}
\ead{kherb@phys.ethz.ch}
\fntext[ETH]{Department of Physics, ETH Z\"urich, Switzerland}

\begin{abstract}
We present an open-source simulation framework for optically detected magnetic resonance, developed in Python. The framework allows users to construct, manipulate, and evolve multipartite quantum systems that consist of spins and electronic levels. We provide an interface for efficient time-evolution in Lindblad form as well as a framework for facilitating simulation of spatial and generalized stochastic dynamics. Further, symbolic operator construction and propagation is supported for simple model systems making the framework also ideal for use in classroom instruction of magnetic resonance. Designed to be backend-agnostic, the library leverages existing Python libraries as computational backends. We introduce the most important functionality and illustrate the syntax on a series of examples. These  include systems such as the nitrogen-vacancy center in diamond and photo-generated spin-correlated radical pairs for which our library  offers system-specific sub-modules.
\end{abstract}

\end{frontmatter}

\flushbottom

\section*{Introduction}

Optically addressable spins are an emerging platform for quantum sensing, communication, and computing. All-optical spin state initialization and readout, together with coherent spin state manipulation, fulfill the majority of the well-known DiVincenzo criteria for quantum information applications. \cite{Divincenzo,Degen2017}
One important class of optically active spins are point defects in wide band gap semiconductors, e.g., color centers in diamond and silicon carbide. The most well-understood defect in diamond, the nitrogen-vacancy (NV) center \cite{Janitz2022}, has been investigated for decades, and along with rapidly growing interest in alternative defects \cite{Zhang2023, Pingault2017} demonstrates significant potential for applications in nanoscale sensing of magnetic\cite{Palm2022} and electric fields\cite{Dolde2011} and for long-lived quantum memories.\cite{Maurer2012} Spin defects hosted in two-dimensional materials such as hexagonal boron nitride offer additional benefits, including simplified on-chip integration and improved light outcoupling \cite{Stern2022}.
Another important class is composed of molecular-based qubit systems, such as photogenerated spin-correlated radical pairs (SCRPs) and optically addressable organic or organometallic motifs \cite{Bayliss2020, Wasielewski2020, Mani2022}. In this regard, chemical synthesis offers atomistic control and great versatility over molecular structure, thereby enabling precise tuning of important characteristics of the quantum system.

The key characteristic of these systems is their spin-dependent photoluminescence, which enables their characterization with optically detected magnetic resonance (ODMR), a highly sensitive technique capable of single-spin detection at room temperature.\cite{Wrachtrup1993}  Simulation of ODMR experiments and, more generally, the convoluted spin and optical dynamics of such systems can be challenging. Accurate modeling requires incorporation of both the coherent spin dynamics as well as incoherent optical excitation and decay dynamics between the electronic states of the system. Established simulation libraries widely used for prediction of conventional nuclear and electron paramagnetic resonance spectra \cite{spinach, easyspin, spindynamica, simpson, gamma} are not well suited for simulations of ODMR experiments and do not utilize free and open-source high-level programming languages. More recently, with a focus on radicals and radical pairs in chemical systems, a C-based toolkit \cite{molspin} has been developed for molecular spin dynamics simulations. Besides these, the physics community has established a number of simulation frameworks in the field of quantum optics \cite{quopticsjl, qutip} and quantum computing \cite{qiskit}. While these platforms feature advanced algorithms for time evolution of open quantum systems, they are not optimized for facile handling of generalized, multipartite systems. \\

Here, we introduce \simos, a simulation library for optically addressable spins in Python. \simos provides a simple yet versatile interface for the construction, manipulation, and time-evolution of operators and superoperators for quantum systems of spins and electronic levels. The time evolution under combined coherent and incoherent dynamics is formulated in Lindblad form \cite{Lindblad1976, Manzano2020} and solver routines facilitate fast implementation and study of complex protocols. Further, spatial dynamics and stochastic modulation of system parameters may be incorporated using a flexible Fokker-Planck framework \cite{Kuprov2016}. As the computational backend, \simos leverages established Python libraries (see below). To further offload the computational workload, an interface is provided for parallel time integration using batched BLAS (Basic Linear Algebra Subprograms) routines on graphical processing units (GPUs) introduced recently by some of us \cite{parament}.
\pagebreak

This overview paper is structured as follows: After introducing the backend-agnostic concept of \simos (Section \ref{sec:backends}), we provide a short description of the basic syntax and functionality that enables handling of multipartite quantum systems in Section \ref{sec:systemconstruction}. We continue with an introduction to the dynamic evolution of quantum systems under coherent and incoherent interactions (Section \ref{sec:introdynamics}) and outline numerical methods for time-dependent and stochastic dynamics generators (Section \ref{sec:numerics}). Sections \ref{sec:2spinsystem}--\ref{sec:fokkerexample} illustrate the syntax and capabilities of the \simos library by modeling a series of selected examples including prototypical two-spin systems, ODMR with the NV center, chirality and magnetic field effects in SCRPs, and heteronuclear decoupling by magic angle spinning.

\section{Interfaced Python Libraries}
\label{sec:backends}

\begin{figure}
\centering
\includegraphics[width=0.5\textwidth]{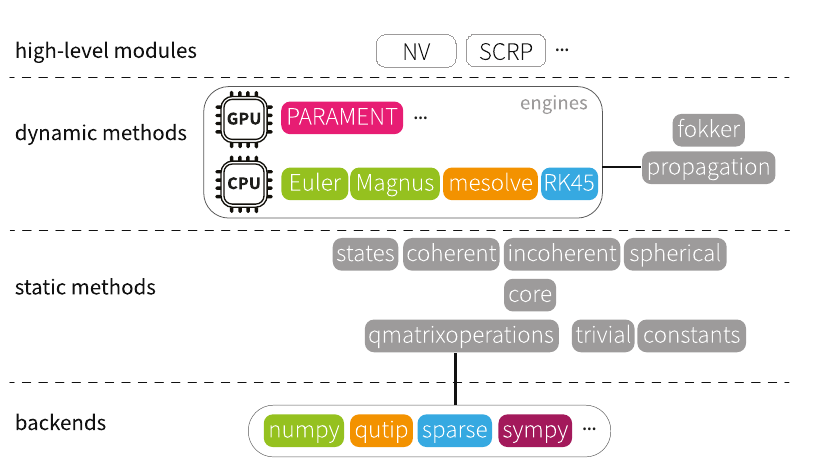}
\caption{Schematic overview of the code hierarchy of \simos. Distinct python libraries serve as computational \emph{backends}, \emph{i.e.} determine the data type of quantum objects and provide implementations for basic matrix operations (\texttt{qmatrixoperations}). \emph{Static methods} are implemented in a backend-agnostic manner, which allows state initialization and interactions of quantum system. Their time evolution can be calculated with a series of \emph{dynamic methods} that outsource the the computation to various  CPU or GPU based \emph{engines}. Finally, \emph{high-level models} for commonly studied systems provide ready-to-use and state-of-the-art implementations.}
\label{fig:simosoverview}
 \end{figure}

 Python is a modern high-level programming language that is free and open-source and therefore widely used in scientific education and original research. 
 To leverage recent efforts of the Python, spin physics, and spin chemistry communities, \simos is implemented in a backend-agnostic manner such that existing Python packages may be used as computational backends. Currently, four third-party libraries are readily available:
 \begin{itemize}
     \item The \emph{QuTiP} \cite{qutip} backend ensures compatibility with the \emph{QuTiP} library, such that \simos users benefit from all existing \emph{QuTiP}-functionality.
     \item The \emph{NumPy} \cite{numpy} backend offers high computational performance and enables portability (e.g. for utilization of \simos on computer clusters).
     \item The \emph{SciPy.sparse} \cite{scipy} backend optimizes computational performance for large yet sparse systems.  
     \item The \emph{SymPy} \cite{sympy} backend enables symbolic operator construction and time propagation often needed for understanding small prototype systems. It is not suited for extensive simulations of large systems but especially useful for educational purposes or conceptualization. 
\end{itemize}
The backend is selected in \simos routines \textit{via} the \texttt{method} keyword argument so that the same code can be run on different backends with only minor modifications. Figure \ref{fig:simosoverview} schematically visualizes the code hierarchy of \simos and involvement of backends. The backend determines the data type of all quantum objects and provides implementations for basic matrix operations of quantum mechanics. These are utilized by static and dynamic methods and ultimately, in high-level and system-specific modules.

\section{Multipartite Quantum Systems}
\label{sec:systemconstruction}
Constructing and manipulating operators of multipartite quantum systems in a convenient manner is the key functionality of \simos.

\begin{figure*}[t]
\centering
\includegraphics[width=1.0\textwidth]{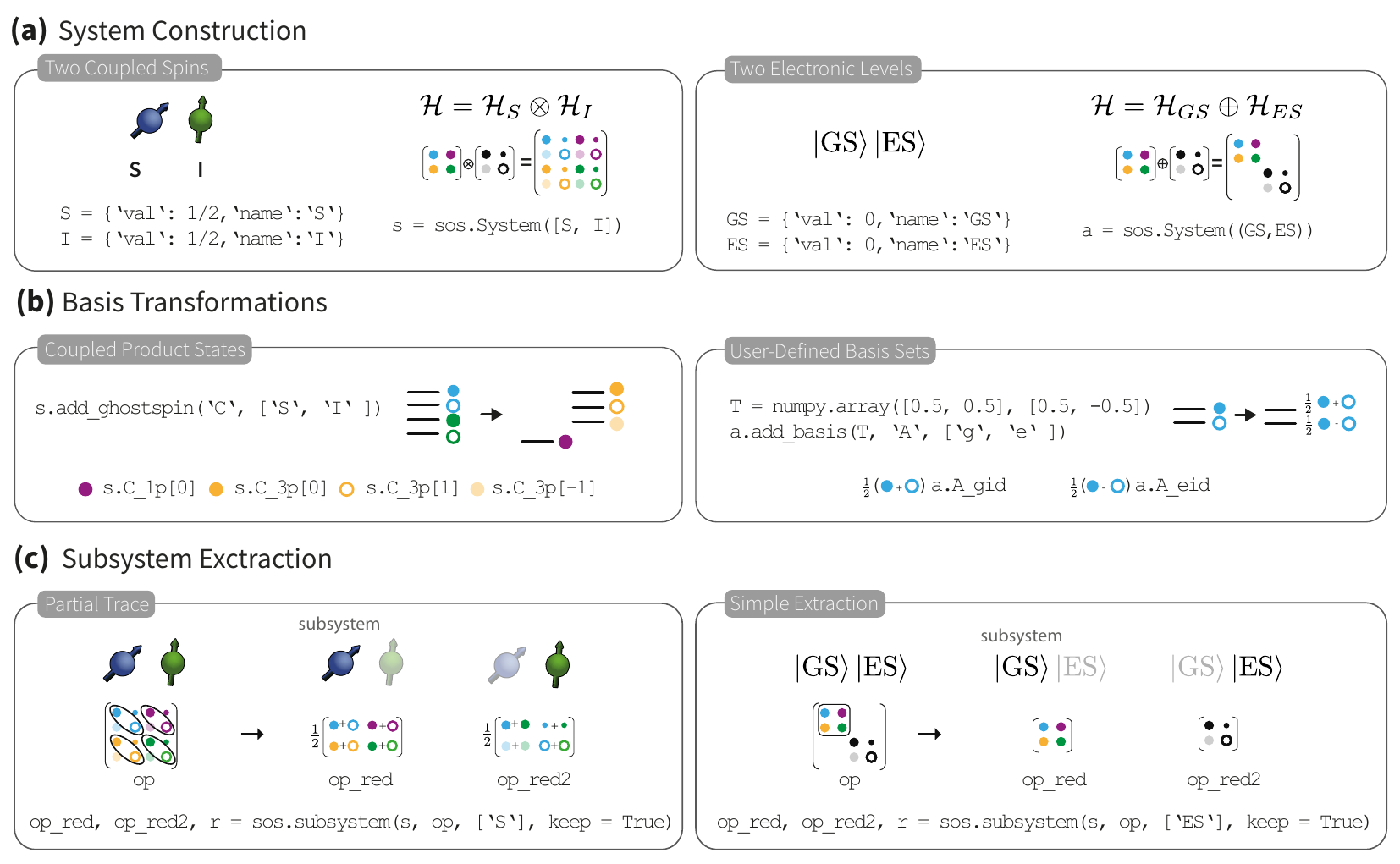}
\caption{Handling multipartite quantum systems in \simos, including Python syntax examples.   \textbf{(a)} The construction of composite Hilbert spaces using tensor products ($\otimes$) and direct sums ($\oplus$) for two examples: a pair of coupled spins $S$ and $I$ (\textit{left}) and a pair of electronic levels $\ket{\mathrm{GS}}$ and $\ket{\mathrm{ES}}$ (\textit{right}). The Hilbert space construction is visualized with symbolic matrices, encoding matrix elements of the first spin (level) in color and the matrix elements of the second spin (level) as shapes. In the combined Hilbert space, products of these matrix elements are encoded in the combined color-shape appearance of the symbols. \textbf{(b)}  \textit{Left}: Coupled product states of spins may be introduced with the \texttt{add\_ghostspin} method. Here, spins $S$ and $I$ are represented as a ghost spin $C$ with singlet  \texttt{C\_1} and triplet \texttt{C\_3} contributions. The underlying transformation from a Zeeman to a coupled singlet-triplet basis is visualized with a level scheme, where total spin number (color) and magnetic spin number (shade) is encoded schematically.  \textit{Right}: User-defined alternate basis states can be introduced with the \texttt{add\_basis} method. We visualize the coupling of two electronic levels ($\ket{\mathrm{GS}}$ and $\ket{\mathrm{ES}}$) to superposition states ($g$ and $e$) that together compose a basis $A$. 
\textbf{(c)} Sub-parts of the multipartite system can be extracted with the \texttt{subsystem} routine, which extracts subsystems using partial traces (\textit{left}) or a simple extraction of specific matrix subsets (\textit{right}). Encircled groups of matrix elements illustrate the extraction of the desired subsystem (\texttt{op\_red}) from the full system operator (\texttt{op}). Matrix representation of both, the desired subsystem (\texttt{op\_red}) as well as the remaining part (\texttt{op\_red2})), are schematically visualized. }
\label{fig:systemconstruction}
\end{figure*}

\subsection{System Construction}

Any spin dynamics simulation requires the construction of system operators in a suitable basis, spanning the full Hilbert space of the multipartite system. Arbitrarily complicated quantum systems may be constructed from the Hilbert spaces of the individual system components with only two mathematical operations, (i) a tensor product and (ii) a direct sum. In \simos, the quantum system is initialized as an instance of the \texttt{System} class, which holds all system operators pre-built upon initialization.  The class constructor is called with a \texttt{systemarray} specifying all members of the quantum system and how their individual Hilbert spaces are combined. Each member of the spin system is defined as a Python dictionary with keys for its name and spin value. In the  \texttt{systemarray} all dictionaries are combined in a series of nested lists and tuples. Lists (indicated by square brackets) indicate that Hilbert spaces are combined using tensor products while tuples (indicated by round brackets) indicate combination with a direct sum. Figure \ref{fig:systemconstruction} (a) illustrates this syntax on two  basic examples, (i) a coupled spin pair and (ii) a pair of electronic levels. 

The system class instance (in our examples typically denoted as \texttt{s}) holds identity operators for all members of the quantum system. For example, \texttt{s.Aid} is used for a member with name \texttt{A}) combined with x, y, z, lowering, raising and projection operators for all spins. These are constructed as \texttt{s.Ax}, \texttt{s.Ay}, \texttt{s.Az}, \texttt{s.Aplus}, \texttt{s.Aminus}, \texttt{s.Ap[0.5]} and \texttt{s.Ap[-0.5]} for a spin $\nicefrac{1}{2}$ with name \texttt{A}). These operators provide the basis for further construction of initial state vectors and density matrices, the system Hamiltonian, collapse operators and superoperators.

\subsection{Alternative Basis Sets and Basis Transformations}

Upon system construction, the spin operators are initialized in the Zeeman basis, spanned by the magnetic quantum numbers of the individual spin members. To include alternative basis sets and transform operators between various bases, the \texttt{System} class provides users with two specific methods.  

Coupled product states of pairs or groups of spins are useful representations for systems in which spin-spin interactions dominate the system Hamiltonian. In \simos, they can be constructed with the \texttt{add$\_$ghostspin} method of the \texttt{System}. The simplest example, the coupled representation of two coupled spin $\nicefrac{1}{2}$ particles as a singlet ($S_{\mathrm{tot}} = \nicefrac{1}{2}-\nicefrac{1}{2} = 0$) and triplet ($S_{\mathrm{tot}}  = \nicefrac{1}{2}+\nicefrac{1}{2} = 1$) is illustrated in Figure \ref{fig:systemconstruction} (b). Here, spins $S$ and $I$ are coupled to obtain the singlet and triplet  \textit{ghostspins}  $C\_1$ and $C\_3$. The operators of the \textit{ghostspins} are generated in full analogy to the `native' spins of the system. For example, the projector onto the the $m_S=0$ level of the triplet is obtained as \texttt{C$\_$3p[0]}. The matrix representations of these operators are still formulated in the Zeeman basis. However, the \texttt{add$\_$ghostspin} method also constructs the transformation matrices to transform operators to or from the coupled basis and stores them as attributes of the \texttt{System}. In our simple example, the transformation matrices can be assessed as \texttt{s.toC} and \texttt{s.fromC}. Using these matrices, any system operator may be transformed to or from the Zeeman to the coupled singlet-triplet basis. For spin systems with more constituents, the \texttt{add$\_$ghostspin} routine enables an arbitrary number of spins to be coupled in a user-defined order.

A completely user-defined basis may be defined using a second method, \texttt{add$\_$basis}, by providing (i) a transformation matrix from the Zeeman basis to the new basis and (ii) a name for the basis and a list of names for all basis states. In Figure \ref{fig:systemconstruction} (b) this method is illustrated for the example of a pair of electronic levels. The method creates identity operators of all new basis states as well as transformation matrices for back and forth conversion between the Zeeman and the alternate basis.

\subsection{Subsystems}

The extraction of parts of a multipartite quantum system (in the following referred to as subsystems) is an essential task during simulation routines and is supported in \simos. For composite systems whose combined Hilbert space was constructed with a tensor product, the subsystem is extracted with a partial trace. If the composite system was constructed with a direct sum, subsystems are direct subsets of the full system operator by means of projection. In Figure \ref{fig:systemconstruction} (c) the extraction of subsystems is illustrated for the  examples introduced in part (a). For larger multipartite systems, our algorithm extracts the desired subsystem in a top-down approach. This routine does not remove members unless they are fully separable from the desired subsystem.

\section{Time Evolution under Coherent and Incoherent Dynamics}
\label{sec:introdynamics}

Once a suitable operator basis has been constructed, generators for quantum dynamics may be built and utilized to simulate the time evolution of the quantum system. 

\subsection{Coherent Evolution in Hilbert Space}
The time evolution of closed quantum systems is governed by coherent interactions which can be formulated as Hamiltonian operators $H$ in the Hilbert space of the quantum system. If the system is in a pure state, represented by a state vector $\ket{\psi}$, the time evolution is described by the Schr\"odinger equation
\begin{equation}
\frac{\partial}{\partial t} \ket{\psi(t)} = -\frac{i}{\hbar} H \ket{\psi(t)} = -i \mathcal{H} \ket{\psi(t)}.
\label{eq:schr}
\end{equation}
ODMR experiments are often conducted on statistical ensembles of quantum systems, characterized by probability distributions of a series of state vectors $\{\ket{\psi_i}, p_i \}$. These mixed states are then represented by density matrices $\rho = \sum_i p_i \ket{\psi_i}\bra{\psi_i}$ and the time evolution is described by the Liouville-von-Neumann equation
\begin{equation}
\frac{\partial}{\partial t} \rho(t) = -i [\mathcal{H}, \rho(t)].
\label{eq:lvn}
\end{equation}

If the Hamiltonian $\mathcal{H}$ is stationary in time, the exact solution of the Schr\"odinger or the Liouville-von-Neumann equation is given by
\begin{subequations}
        \begin{equation}
            \ket{\psi(t)} = e^{-\mathrm{i} \mathcal{H} t} \ket{\psi(t=0)} = U(t) \ket{\psi(t=0)} \quad\text{and}
        \end{equation}
        \begin{equation}
            \rho(t) = U(t) \, \rho(0)\, U^\dagger(t) \text{,}
        \end{equation}
    \end{subequations}
respectively.
In \simos,  the propagator $U$ as well as the evolved state vector $\ket{\psi(t)}$ or density matrix $\rho(t)$ can be computed using the \texttt{evol} routine. 
The evolution under a time-dependent Hamiltonian uses a separate routine that is introduced in detail in Section \ref{sec:numerics}. 

\begin{figure*}[t]
\centering
\includegraphics[width=1.0\textwidth]{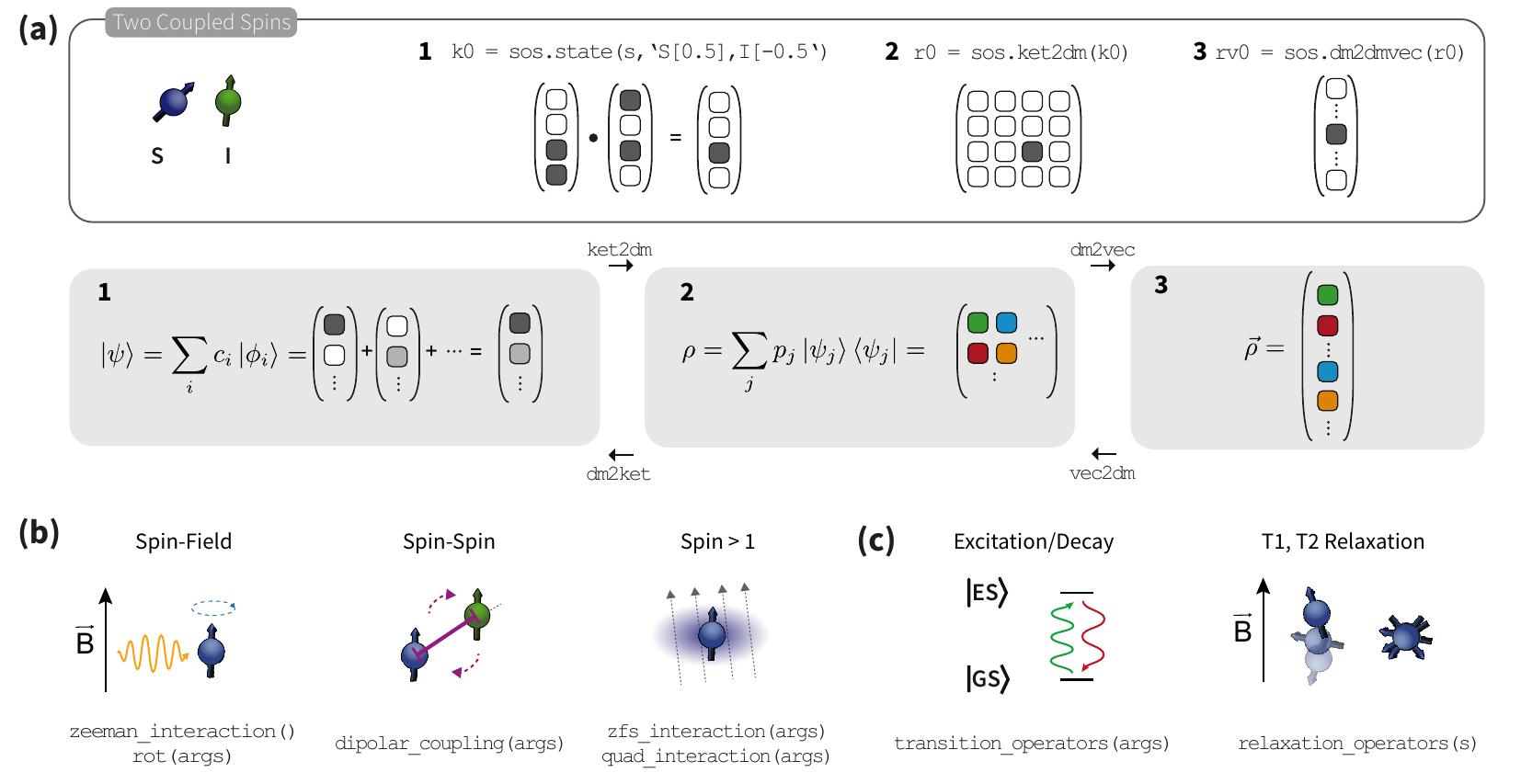}
\caption{Describing states and interactions of quantum systems in \simos. (a) \textit{Top:} State construction for a two-spin system.  If the system is in a pure state, e.g., spin $S$ is polarized up ($m_S = 0.5$) and spin $I$ is polarized down ($m_S = -0.5$), a state vector $\ket{\psi}$ may be used to describe the initial state. Here, we demonstrate how to utilize the \texttt{state} function of \simos to initialize the state vector \texttt{k0} (1) and its transformation into a density matrix \texttt{r0} (2) and vectorized density matrix \texttt{rv0} (3). \textit{Bottom}: Generalized visualization of state vectors (1), density matrices (2) and vectorized density matrices (3). \simos methods for interconversion between the individual representations, i.e., \texttt{ket2dm} and \texttt{dm2ket} for 1 to 2 and \texttt{dm2vec} and \texttt{vec2dm} for 2 to 3, are indicated.  (b) Hamiltonian operators for common coherent interactions are readily available in \simos. These include (from left to right) spin-field interactions (\texttt{zeeman\_interaction}) , spin-spin couplings (\texttt{dipolar\_coupling}) as well as zero-field splittings or quadrupole interactions (\texttt{zfs\_interaction}, \texttt{quad\_interaction}) for spins $S > 1$. (c) Incoherent interactions require the generation of collapse and/or jump operators. For photoinduced dynamics, the \texttt{transition\_operators} routine can be used to construct collapse operators by providing a suitable dictionary. For phenomenological spin relaxation, the \texttt{relaxation\_operators} routine can be used. The method utilizes phenomenological $T_1$ and $T_2$ values that must be specified in the attributes dictionary of individual system members.}
\label{fig:statesandinteractions}
\end{figure*}

Both the initial system state $\rho(0)$ and the system Hamiltonian $\mathcal{H}$ can be generated using either the operators of a \textit{System} class instance or by making use of further \simos functionality as outlined in the following. The generation of initial states in \simos is visualized for a two-spin system in Figure \ref{fig:statesandinteractions} (a). The \texttt{state} function initializes pure state vectors and conversion between state vector and density matrices if performed with the \texttt{ket2dm} and \texttt{dm2ket} functions.  Additional methods initialize polarized or thermal states and are detailed in our online documentation.  

Methods to facilitate the construction of Hamiltonian operators target the most common sources of coherent dynamics of optically addressable spins. As visualized in Figure \ref{fig:statesandinteractions} (b) these include interactions of spins with magnetic fields (\texttt{zeeman\_interaction}),  pairwise spin-spin interactions (\texttt{dipolar\_coupling}) as well as interactions of spins $ S > 1$  (\texttt{zfs\_interaction}, \texttt{quad\_interaction}).
Since all coherent spin-field or spin-spin interactions share a universal form 
\begin{equation}
\mathcal{H} =  \vec{S} \cdot \mathbf{A} \cdot \vec{X},    
\label{eq:Hmat}
\end{equation}
where $\vec{S}$ is a vector of Cartesian spin operators, $\textbf{A}$ is a 3$\times$3 matrix and $\vec{X}$ is either a spin-operator vector for a second spin,  or a magnetic field vector, we further provide the general routines \texttt{interaction\_hamiltonian} and \texttt{AnisotropicCoupling}. While \texttt{AnisotropicCoupling} facilitates conversion between different conventions for the interaction matrix $\textbf{A}$, \texttt{interaction\_hamiltonian} 
is used to construct any Hamiltonian operator fulfilling  \eqref{eq:Hmat}. Notably, \simos also provides a comprehensive look-up table of gyromagnetic ratios and other physical constants. In principle, optical transitions can also become part of the coherent system dynamics \cite{Chu2014}. Although this regime was not the main focus during development, the respective operators can be constructed utilizing the complete quantum mechanical basis that is initialized upon system construction.

\subsection{Incoherent Evolution in Liouville Space}

Systems that exhibit spin-dependent photoluminescence are typically open quantum systems. Their incoherent interaction with the environment interferes with their coherent time evolution and an accurate simulation of their dynamics requires a quantum master equation (QME). Since a QME enables a non-unitary time evolution of quantum states, a description with density matrices becomes mandatory and the combined coherent and incoherent system dynamics are formulated as superoperators in Liouville space. Multiple approximate QMEs exist and are preferentially utilized among specific scientific communities.  However not all of them generate completely positive time-evolution operators that preserve the trace of the density matrix ($tr(\rho(t))=1$). Importantly, if a QME does not generate a completely positive trace preserving (CPTP) map, it may produce solutions that are not physical and therefore require further correction. 

The Lindblad theorem \cite{Lindblad1976} states that the generator of any quantum operation that satisfies the CPTP criterion can be written in the form 

\begin{equation}
\frac{d}{dt} \rho =  -i [\mathcal{H}, \rho] + \sum_k \left(\mathcal{L}_k\rho\mathcal{L}_k^{\dagger} - \frac{1}{2} \{\mathcal{L}_k^{\dagger}\mathcal{L}_k \rho\} \right) 
\label{eq:QMELindblad}
\end{equation}

where $[\cdot , \cdot]$ and $\{\cdot , \cdot\}$ are the commutator and the anticommutator, $\mathcal{H}$ is the system Hamiltonian and $\mathcal{L}_k$ are collapse (or jump) operators. The microscopic derivation of the Lindblad form involves several assumptions, most importantly that the dynamics are Markovian and that the system is weakly damped. Under these assumptions the density matrix of the system remains separable from the bath's degrees of freedom at all times. An extensive discussion of the Lindblad form of the QME may be found elsewhere  \cite{Campaioli2024}. Here, we only discuss the most important characteristics that are necessary to grasp the implementation in \simos.

The first part of \eqref{eq:QMELindblad} is the Liouville--von Neumann equation that describes the coherent dynamics discussed above. Without incoherent contributions,  \eqref{eq:QMELindblad} simplifies accordingly and can be propagated in Hilbert space.
The second part is a dissipative superoperator, which describes incoherent, stochastic state transitions characterized by jump operators and transition rates.

Common sources of incoherent dynamics include
\begin{itemize}
\item Optical excitation and decay of electronic transitions at non-cryogenic temperatures,
\item Dissipative interaction with a quantum mechanical environment (bath),
\item Stochastic modulations of (classical) system parameters (e.g., rotational diffusion in a liquid environment, flow in a field gradient, static field drifts).    
\end{itemize}

The first source of incoherent dynamics, optical excitation and decay events, are characterized by collapse operators of the type $\ket{m}\bra{n}$ for pairs of electronic levels $m, n$ and classical transition rates that are available for many systems. In \simos, the \texttt{transition\_operators} function automatically generates the respective collapse operators from a user-defined rate dictionary as the input argument. Figure \ref{fig:statesandinteractions}(c) illustrates the incoherent excitation and decay of a simple two-level system. In this case, the rate dictionary can be defined as

\begin{lstlisting}[language=Python]
rates = {}
kex = 5e6 # excitation rate in Hz
kdec = 10e6 # decay rate in Hz
rates["GS -> ES"] = kex
rates["ES -> GS"] = kdec
c_ops = sos.transition_operators(a, rates)
\end{lstlisting}

where the collapse operators of the dynamics of system \texttt{a} can be obtained with only a few line of code.

For the other two sources of incoherent dynamics, the construction of suitable collapse operators is usually more complicated. In both cases, construction of effective collapse operators requires knowledge about the time correlation functions of the bath and the spectral density of the system. The underlying theory was originally developed for the case of a true quantum mechanical bath and later adapted for the semi-classical case. In conventional EPR and NMR spectroscopy, semi-classical Bloch-Wangsness-Redfield theory (BWR) is most commonly used to construct relaxation superoperators. Their purpose is to account for stochastic modulation of system parameters due to spatial dynamics of anisotropic systems.\cite{Rodin2022} However, BWR theory performs poorly for systems far from equilibrium  and is therefore ill-suited for simulations of optically active spins whose initial states are typically non-Boltzmann populations.\cite{Bengs2020}

In \simos, phenomenological longitudinal ($T_1$) and transverse ($T_2$) relaxation may be included using the \texttt{relaxation\_operators} function which returns a set of collapse operators for use with our time propagation routines. Rate constants are added to the system's dictionary using e.g., 
\begin{lstlisting}[language=Python]
S = {'name':'S', 'val':1/2,'T1':1e-3}
\end{lstlisting}

If the relaxation is induced by the aforementioned stochastic modulation of system parameters, most commonly spatial dynamics, users may alternatively use the Fokker--Planck framework to explicitly simulate stochastic parameter modulation (Section \ref{sec:numerics}).

Equation \ref{eq:QMELindblad} can be formulated in Liouville space as
\begin{subequations}
    \begin{equation}
        \frac{\partial}{\partial t} \vec{\rho}  = \mathbf{L}\vec{\rho} = (\mathbf{H} + \mathbf{G}) \vec{\rho} 
    \end{equation}
where $\mathbf{L}$ is the Lindbladian superoperator which acts on a vectorized density matrix, $\vec{\rho}$. The vectorization of density matrices is illustrated schematically in Figure \ref{fig:statesandinteractions} (a) and obtained in \simos with the \texttt{dm2vec} and \texttt{vec2dm} methods. The Lindbladian may be separated into a coherent (Hamiltonian) part 
\begin{equation}
    \mathbf{H} = -\mathrm{i} (\mathcal{H} \otimes \mathbbm{1} - \mathbbm{1} \otimes \mathcal{H})
\end{equation}
and an incoherent part
\begin{equation}
    \mathbf{G} = \sum_k \mathcal{L}^\dagger_k \otimes \mathcal{L}_k -\frac{1}{2} \mathbbm{1} \otimes  (\mathcal{L}_k^\dagger \mathcal{L}_k) - \frac{1}{2}  (\mathcal{L}_k^\dagger \mathcal{L}_k)  \otimes \mathbbm{1}.
\end{equation}
\end{subequations}
Importantly, \eqref{eq:QMELindblad} is a
linear equation in Liouville space. Thus, the solution is again obtained as
\begin{equation}
\vec{\rho}(t)  = e^{(\mathbf{H}+\mathbf{G}) t}\, \vec{\rho} (t=0),
\end{equation}
\textit{via} the calculation of matrix exponentials.

The \texttt{evol} routine in \simos supports the inclusion of incoherent dynamics and accepts a list of collapse operators $\mathcal{L}_k$ as the \texttt{c$\_$ops=[...]} keyword argument. Although the actual propagation is performed in Liouville space, users do not have to perform the vectorization and superoperator construction themselves and may simply provide and retrieve Hamiltonian and density operators in their Hilbert space representations.

\section{Numerical Methods for Time-Dependent and Stochastic Generators}
\label{sec:numerics}

\begin{table*}[t]
    \centering
    \caption{Overview of propagation engines for time-dependent Hamiltonians}
    \begin{tabular}{llll}
    \textbf{Engline} & \textbf{Description} & \textbf{Characteristics} &  \textbf{Parallelizable} \\
    \hline
        \texttt{cpu} & CPU-based Euler-type & \tabitem Easy to follow & Yes
        \vspace{0.75em}\\
        \texttt{cpu} with \texttt{magnus=True} & CPU-based Magnus integrator & \tabitem Portabilty\\
        & & \tabitem Decent speed\\
        & & \tabitem Good convergence & Yes  
        \vspace{0.75em}\\
        \texttt{parament} & GPU-based Magnus integrator & \tabitem Speed \\
         & & \tabitem Good convergence     & Yes (inherent) 
       \vspace{0.75em}\\
        \texttt{qutip} & QuTiP \texttt{mesolve} & \tabitem Established \& battle-field tested  & No
        \vspace{0.75em}\\
        \texttt{RK45} & \texttt{scipy.integrate} 4th order Runge-Kutta & \tabitem Memory saving  & No\\ 
        \end{tabular}
    
    \label{tab:engines}
\end{table*}
The coherent and incoherent interactions introduced in Section \ref{sec:introdynamics} may be time-varying or stochastically modulated. \simos provides methods that facilitate the simulation of quantum systems under non-static Hamiltonians (Liouvillians).  

\subsection{Time-Dependent Generators}

Hamiltonian and collapse operators may become time-dependent under the application of time-varying control fields, e.g., magnetic or electric fields or shaped laser-excitation pulses. In most cases, the dynamic behavior can be efficiently parametrized with a limited set of control fields and collapse operators without any loss of generality. The time-varying Hamiltonian and collapse operators are formulated as
\begin{equation}
    \mathcal{H}(t) = \mathcal{H}_0 + \sum_{i=1}^N c_i(t) \mathcal{H}_i
\end{equation}
and
\begin{equation}
    \mathcal{L}_k(t) = \mathcal{L}_{k,0} + \sum_{i=1}^N C_i(t) \mathcal{L}_{k,i}
\end{equation}
with time-independent basis functions  $\mathcal{H}_i$ and $\mathcal{L}_{k,i}$ and control amplitudes $c_i(t)$ and $C_i(t)$. 

The \texttt{prop} routine of \simos accepts the time-independent parts $\mathcal{H}_0$ ($\mathcal{L}_{k,0}$) as well as lists of time-varying control operators $\mathcal{H}_i$ ($\mathcal{L}_{k,i}$) and their control amplitudes $c_i(t)$ ($C_i(t)$) as input arguments. The time-dependence of the control amplitudes is assumed to be piecewise constant and control amplitudes are specified for discretized time intervals \texttt{dt}.  This functionality is only compatible for our numerical backends and not for our symbolic \texttt{sympy} backend.

\subsection{Engines}

The development of computationally efficient integration schemes for the numerical propagation of quantum systems is an active field of research.  Besides classical ordinary differential equation (ODE) solvers, Euler-type integrators are common in the field of magnetic resonance simulations. Here, in full analogy to the static solutions presented in Section \ref{sec:introdynamics}, the matrix exponential of the piecewise constant Hamiltonian or Liouvillian is evaluated separately for each time interval. The repeated computation of the matrix exponential, however, is computationally expensive. A more efficient implementation is the use of parallel-in-time integrators such as PARAMENT \cite{parament}, developed by some of the authors. We have further shown that Euler-type integrators can be easily converted into Magnus-type integrators that benefit from better convergence \cite{parament,magnus}. 

\simos' \texttt{prop} routine serves as an interface to different so-called \emph{engines} that perform the calculation and are readily selected \textit{via} the \texttt{engine} keyword argument. Users can thereby select an integrator scheme that is optimal for their specific problem characteristics such as dimensionality and sparsity of operators. Table \ref{tab:engines} provides an overview of the available engines. The \texttt{cpu} and \texttt{parament} engines are CPU- and GPU-based Magnus-type integrators using the PARAMENT package. The \texttt{parament} engine directly interfaces PARAMENT while the \texttt{cpu} engine utilizes the matrix exponential function provided by SciPy. Here, users may further chose between Euler and Magnus-type integrators \textit{via} the \texttt{magnus} option. Besides this, ODE solvers based on the Runge-Kutta method are available through the \texttt{RK45} and \texttt{qutip} engines. The  \texttt{RK45} engine is built on SciPy integrators while the \texttt{qutip} engine uses QuTiP's native \texttt{mesolve} method. The inclusion of native \texttt{qutip} functionality is enabled by the backend-agnostic implementation of \simos. Importantly, ongoing development in QuTiP or other Python libraries is therefore readily available in \simos simulations.

\subsection{Laboratory Frame Simulation Helpers}

The most common source of time-varying dynamics in magnetic resonance simulations are electromagnetic control fields ("pulses") with well-defined frequency, phase and amplitude. To reduce computational cost and complexity, simulations are often performed in a \emph{rotating frame} of reference which is accelerated at a frequency $\omega_\mathrm{rf}$ compared to the stationary frame of reference. The stationary frame of reference is typically called \emph{laboratory frame} in the context of magnetic resonance. %
The time-dependent Hamiltonian may be converted into a stationary Hamiltonian under the rotating wave approximation (also denoted high-field approximation in  NMR). 

However, this approximation is not always applicable and in some cases, laboratory frame simulations that explicitly include the time-dynamics of the control-fields are required. Laboratory frame simulations do not only suffer from enhanced computational cost, but also require proper handling of the phase of the control fields. To facilitate this aspect of laboratory frame simulations, \simos provides the \texttt{Wallclock} object. It can be passed to all functions in SimOS that generate (shaped) control pulses (such as \texttt{square$\_$pulse}, \texttt{arb$\_$pulse} etc.) as well as the \texttt{evol} and \texttt{prop} routines. The Wallclock object keeps track of the passed time and adjusts the phase accordingly, thereby allowing users to off-load the tedious and error-prone book keeping to \simos. By default, a global Wallclock (\texttt{wallclock='global'}) is used, which handles multiple function calls for a single laboratory frame simulation.

\subsection{Stochastic Modulation of System Parameters and Fokker-Planck Formalism}
\begin{figure*}[h!]
    \centering
    \includegraphics[width=0.95\linewidth]{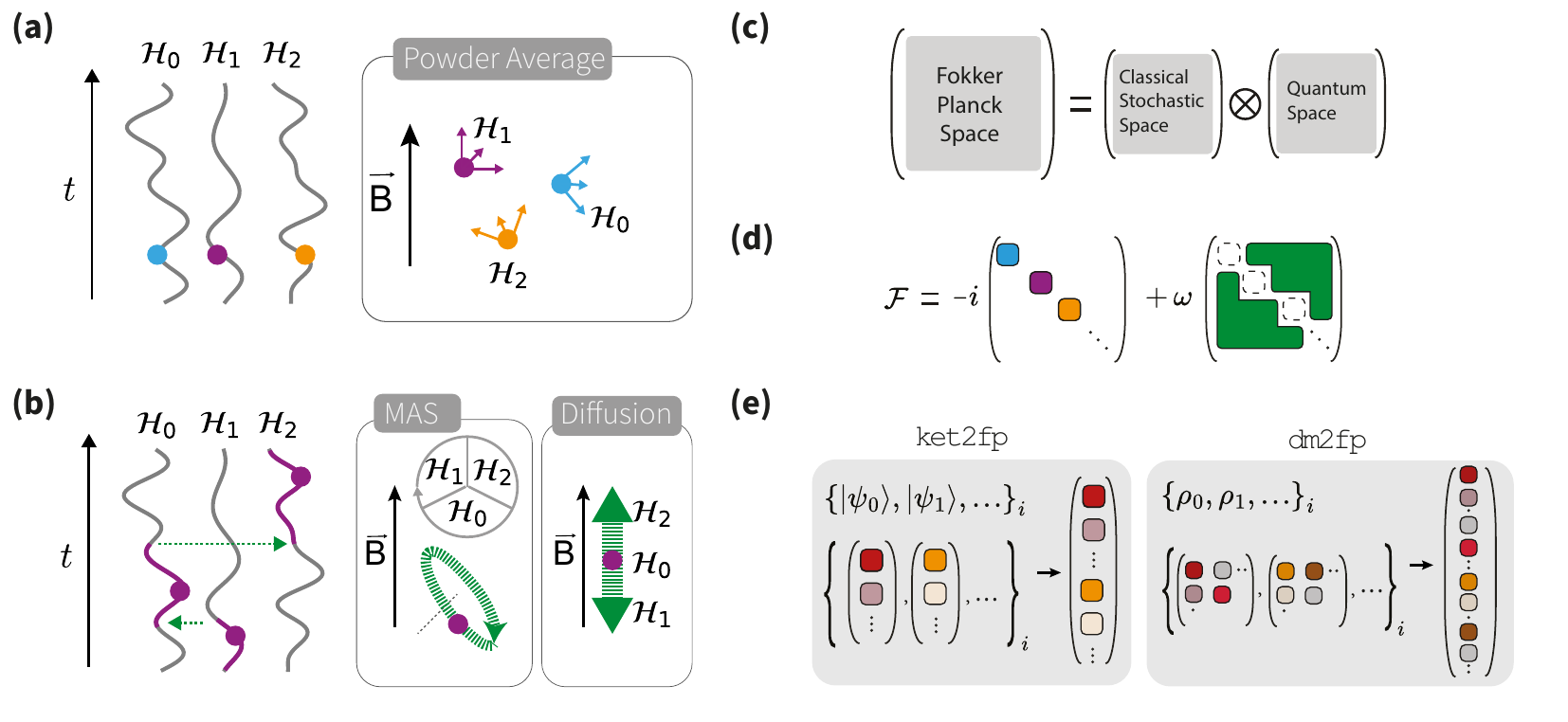}
    \caption{The Fokker--Planck module. (a) \textit{Left}:  Stochastic contributions to the system Hamiltonian are not uniform among all members of the ensemble. Individual ensemble members evolve along different system Hamiltonians (here denoted as  $\mathcal{H}_0, \mathcal{H}_1, \mathcal{H}_2$) and thus different evolution trajectories. \textit{Right}: A common example for a stochastic ensemble is the powder average of  NMR where anisotropic contributions to the system Hamiltonian vary for different molecular orientations with respect to an external magnetic field $\vec{B}$. Each distinct orientation represents an individual system Hamiltonian $\mathcal{H}_i$ and evolution trajectory of the stochastical ensemble.  (b) \textit{Left}: A system may transition between the trajectories during time evolution as visualized schematically by green dashed arrows. \textit{Right}: Common examples for such stochastic dynamics are spatial dynamics, e.g., magic-angle spinning (MAS) or diffusion. For MAS, the sample is physically rotated around an axis which stands at the magic angle relative to the external magnetic field $\vec{B}$. We identify different phases of the rotation as distinct stochastic trajectories (i.e., Hamiltonians, visualized here for an examples of three distinct rotor phases and corresponding Hamiltonians) and system jumps are induced by the rotational motion. In the case of diffusion, the ensemble is spanned by different position of a system in a magnetic field and jumps are induced by the  diffusive motion (linear or rotational tumbling). 
    (c) The Fokker--Planck space is a product space between a classical stochastic space and the quantum space (i.e., Hilbert or Liouville space). (d) Schematic visualization of the mathematical structure of the Fokker--Planck superoperator $\mathcal{F}$. The first part of $\mathcal{F}$ has a block-diagonal structure. Each individual block (colored square) generates the dynamics of an individual evolution trajectory of the stochastical ensemble (i.e., the dynamics visualized in (a)). The second part of $\mathcal{F}$ is an off-diagonal operator and induces transitions between the different evolution trajectories (i.e. the system jumps visualized in (b)). (e) \simos  provides a method \texttt{ket2fp} (\texttt{dm2fp}) to transform an ensemble of state vectors $\{ \ket{\psi_0},\ket{\psi_1} \dots \}$ (density matrices $\{ \rho_0, \rho_1 \dots \}$) from the Hilbert (Liouville) space of the quantum system into a single vector in Fokker--Planck space.}
    \label{fig:fokker_figure}
\end{figure*}
Although the Schr\"odinger, Liouville, and Lindblad equations are powerful tools to simulate the time-evolution of quantum systems, they are not well suited for incorporating stochastic dynamics. Stochastic contributions can arise for ensembles of quantum systems (multiple system copies or measurement repetitions) if the Hamiltonian or Liouvillian is not uniform across the ensemble. In this case, individual ensemble members evolve along different trajectories and the ensemble average differs from the result for a single member. If these trajectories are not dynamically intertwined, the individual systems can be evolved parallel-in-time with subsequent averaging of their results (Figure \ref{fig:fokker_figure} (a)). An example for such a "static" stochastic situation is a powder average in NMR and EPR spectroscopy. Here, we interpret the ensemble as multiple system copies with different spatial orientations relative to a magnetic field. Such an ensemble leads to variations in anisotropic interactions (e.g., anisotropic chemical shifts or dipolar couplings). However, the trajectories can be intertwined if system jumps occur during time evolution (Figure \ref{fig:fokker_figure} (b)). The ensemble members then may no longer be evolved in an independent manner and simulating the ensemble dynamics becomes non-trivial. A common example for this "dynamic" stochastic situation are spatial dynamics in magnetic resonance experiments, for example, magic angle spinning (MAS) or rotational diffusion. 

Recently, Kuprov and coworkers advocated for the Fokker--Planck formalism as a universal and elegant approach to include stochastic contributions in magnetic resonance simulations.\cite{Kuprov2016} \simos uses this framework in a generalized manner, simplifying the incorporation of arbitrary stochastic dynamics in simulation routines. As illustrated in Figure \ref{fig:fokker_figure} (c), the stochastic dynamics are formulated in a higher-dimensional Fokker--Planck space, obtained by forming the tensor product of the original system space (i.e., the Hilbert or Liouville space of the multipartite system) and a classical state space. The latter discretizes the stochastically varying, classical conditions (e.g., molecule orientation, rotor phase, field value of fluctuating fields). The number of included classical basis functions (i.e., the dimensionality of the classical state space) determines how accurate the stochastic dynamics are being captured. The equation of motion in Fokker--Planck space results is then
\begin{equation}
   \frac{\partial}{\partial t} \vec{\rho}_{\mathrm{FP}} = -i \mathcal{F} = -i \mathcal{Q} \vec{\rho}_{\mathrm{FP}} + \omega \left( \left[ \frac{\partial}{\partial \zeta}\right]^n \otimes \mathbbm{1} \right) \vec{\rho}_{\mathrm{FP}} 
   \label{eq:fokkerplanckdynamics}
\end{equation}
where $\mathcal{F}$ is the Fokker--Planck superoperator and and $\vec{\rho}_{\mathrm{FP}}$ is a vectorized, Fokker--Planck space representation of the ensemble state. For an in-depth discussion of \eqref{eq:fokkerplanckdynamics} we refer the reader to the work of Kuprov and others \cite{Kuprov2016}. Here, we limit the discussion to the mathematical structure of the individual elements. The Fokker--Planck superoperator $\mathcal{F}$ (Figure \ref{fig:fokker_figure} (d)) has two contributions: 
\begin{enumerate} 
\item A block diagonal part $\mathcal{Q}$ where every block holds the Hamiltonian or Liouvillian of a distinct classical condition;
\item An off-diagonal part $\omega \left( \left[ \nicefrac{\partial}{\partial \zeta}\right] \otimes \mathbbm{1} \right)$ that exchanges populations between the classical sub-spaces at a frequency $\omega$.
\end{enumerate}
The Fourier differentiation matrix $\left[ \nicefrac{\partial}{\partial \zeta}\right]^n$ with respect to the classical coordinate $\zeta$ may be of order $n=0,1,2$. If $n=0$, no exchange of population occurs between the classical subspace; if $n=1$, the dynamics are linear; if $n=2$, diffusive dynamics are introduced.  The Fokker--Planck space representation of the ensemble state $ \vec{\rho}_{\mathrm{FP}}$ is obtained by concatenating the the state vectors or density operators of individual system members to a single vector (Figure \ref{fig:fokker_figure} (e)).

\simos provides a specialized object, \texttt{simos.StochasticLiouvilleParameters}, to construct all permutations of the classical state space variables. The function \texttt{simos.stochastic\_evol()} serves as an interface for the simulation. It automatically constructs the Fokker--Planck superoperator $\mathcal{F}$ and transforms and extracts the system state to and from the Fokker--Planck space.

\FloatBarrier

\section{Example 1 -- The DEER Experiment}
\label{sec:2spinsystem}

 To illustrate the syntax of \simos, we begin by simulating a very simple double electron-electron resonance (DEER) experiment.

\begin{figure}[h]
    \includegraphics[width=\linewidth]{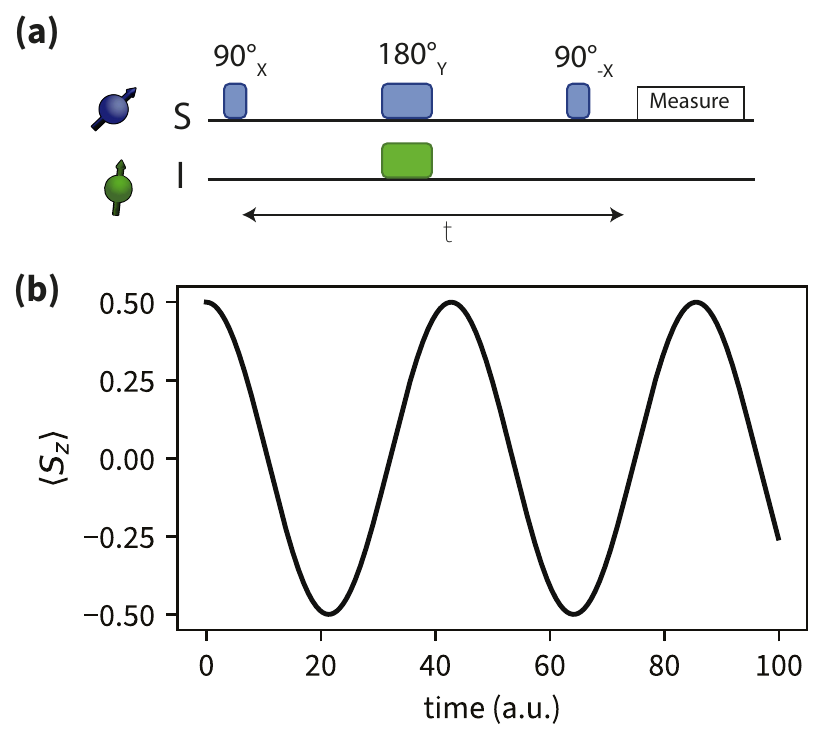}
    \caption{DEER experiment. (a) Schematic visualization of the pulse sequence. The experiment consists of a spin echo on the observer spin (blue) and a single inversion pulse on the pump spin (green). The relative timing $t$ between the central $\pi$ pulse of the echo and the $\pi$ pulse on the pump spin is varied.  (b) The detected echo intensity, $\langle S_z\rangle$ is modulated as a function of the delay time. The strength of the dipolar coupling and therefore the distance between the spins can be estimated from the modulation frequency. }
    \label{fig:deer}
\end{figure}

DEER spectroscopy is a pulsed EPR technique that can be used to measure the distance between two unpaired electrons \textit{via} their dipolar coupling interaction. As shown in Figure \ref{fig:deer} (a), the sequence consists of a spin echo experiment on one spin (observer or probe spin) and a single inversion pulse on the second spin (pump spin). If the echo time $t$ is swept incrementally, the echo intensity is modulated as a function of the dipolar interaction strength. Ultimately, the distance between the two spins can be extracted from the modulation. DEER is thus a powerful tool to gain insight into biomolecule structure, and has been used to study the structure of membrane proteins, protein--protein interactions, and protein--DNA interactions \cite{Jeschke2012}. Other existing variants of the DEER experiment can be simulated similarly.

In Listing \ref{code:deer}, we provide the code snippets for a symbolic DEER simulation together with the mathematical notation of the executed operations. We initialize the system in the SymPy backend (\texttt{method='sympy'}) and define SymPy symbols for all variables. The result of our simulation is a symbolic expectation value
\begin{equation}
    \langle S_z \rangle \propto\cos{\left(\frac{3 t \sin^{2}{\left(\theta \right)}}{8 \pi r^{3}} - \frac{t}{4 \pi r^{3}} \right)}
\end{equation}
for the modulated echo intensity. 
In order to perform the simulation in a numerical manner, the system has to be created with a numeric backend (e.g., \texttt{method='qutip'}) and numerical values must be assigned to all variables. 
Otherwise, the same syntax applies as all core functionality of \simos is completely backend-agnostic.  We utilize a \texttt{for} loop 
\begin{lstlisting}
store = []
for i in range(100):
  psi = sos.rot(s.Sx,np.pi/2,psi0)
  psi = sos.evol(H0,i*dt/2,psi)
  psi = sos.rot(s.Sy,np.pi,psi)
  psi = sos.rot(s.Iy,np.pi,psi)
  psi = sos.evol(H0,i*dt/2,psi)
  psi = sos.rot(s.Sy,np.pi,psi)
  psi = sos.rot(s.Sx,-np.pi/2,psi0)
  m = sos.expect(s.Sz,psi)
  store.append(m)
\end{lstlisting}
 to explicitly evaluate the evolution trajectory which is visualized in Figure \ref{fig:deer}b.

\begin{algorithm*}
\begin{lstlisting}
import sympy as sp
import simos as sos
r,th,phi,y1,y2,t = sp.symbols('r,theta,phi,gamma_1,gamma_2,t', real=True,positive=True)
\end{lstlisting} 
\ContinueLineNumber
\begin{lstlisting}
system_def = []
system_def.append({'name':'S','val':1/2})
system_def.append({'name':'I','val':1/2})
s = sos.System(system_def,'sympy')
\end{lstlisting} 
$H = -\frac{\mu_0\gamma_1\gamma_2\hbar^2}{4\pi r^5 } \left[3 \left({\Vec{S}} \cdot \Vec{r}\right) \left({\Vec{I}} \cdot \Vec{r}\right) - {\Vec{S}} \cdot {\Vec{I}}
\right] \stackrel{\mathrm{sec. approx}}{=}  -\frac{\mu_0\gamma_1\gamma_2\hbar^2}{4\pi r^3 } \left(3 \cos^2\theta - 1\right) S_z I_z$
\ContinueLineNumber
\begin{lstlisting}
H0 = sos.dipolar_coupling(s,'S','I',y1,y2,r,th,phi,approx='secular')
\end{lstlisting} 
$\ket{\psi_0} = \ket{\downarrow}_S \otimes \ket{\downarrow}_I$
\ContinueLineNumber
\begin{lstlisting}
psi0 = sos.state(s,'S[-0.5],I[-0.5]')
\end{lstlisting} 
$\ket{\psi_1} = R_S^{X}(\pi/2) \ket{\psi_0}$
\ContinueLineNumber
\begin{lstlisting}
psi = sos.rot(s.Sx,sp.pi/2,psi0)
\end{lstlisting} 
\ContinueLineNumber
$\ket{\psi_2} = e^{-i {H} t/2} \ket{\psi_1}$
\begin{lstlisting}
psi = sos.evol(H0,t/2,psi)
\end{lstlisting} 
$\ket{\psi_3} = R_S^{Y}(\pi)\,R_I^{Y}(\pi) \ket{\psi_2}$
\ContinueLineNumber
\begin{lstlisting}
psi = sos.rot(s.Sy,sp.pi,psi)
psi = sos.rot(s.Iy,sp.pi,psi)
\end{lstlisting} 
$\ket{\psi_4} = e^{-i {H} t/2} \ket{\psi_3}$
\ContinueLineNumber
\begin{lstlisting}
psi = sos.evol(H0,t/2,psi)
\end{lstlisting} 
$\ket{\psi_5} = R_S^{-X}(\pi/2) \ket{\psi_4}$
\ContinueLineNumber
\begin{lstlisting}
psi = sos.rot(s.Sx,-sp.pi/2,psi)
\end{lstlisting} 
$m = \braket{\psi_5|S_x|\psi_5}$
\ContinueLineNumber
\begin{lstlisting}
m = sos.expect(s.Sz,psi)
\end{lstlisting}
\caption{DEER simulation}
\label{code:deer}
\end{algorithm*}

\section{Example 2:  Nitrogen-Vacancy Centers in Diamond}
\label{sec:examplesNV}

\begin{figure*}[]
\centering
\includegraphics[width=1.0\textwidth]{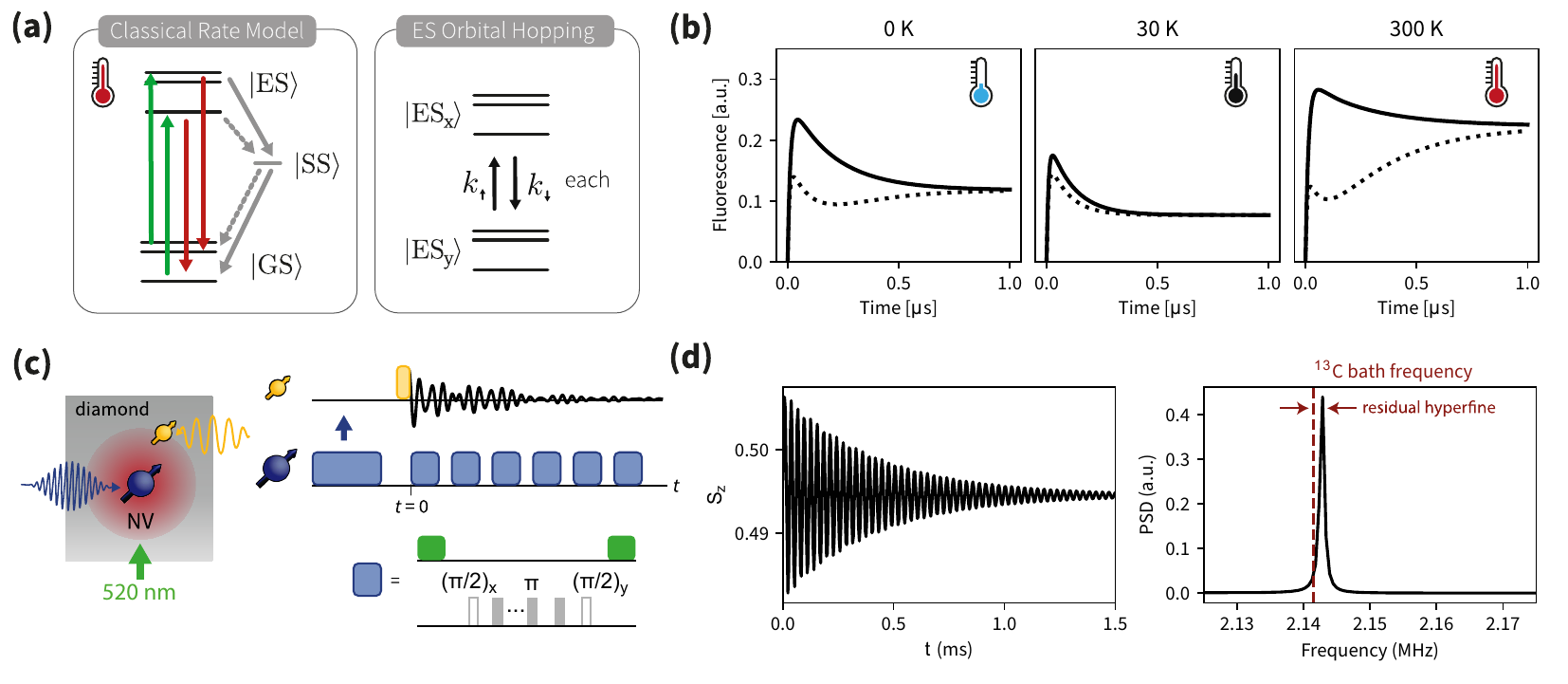}
\caption{Photo-physics (a, b) and nanoscale magnetic resonance (c,d) of the nitrogen-vacancy (NV)  center in diamond. (a) \textit{Left}: Schematic visualization of the classical rate model of the NV center with electronic ground $\ket{\mathrm{GS}}$, excited $\ket{\mathrm{ES}}$ and shelving $\ket{\mathrm{SS}}$ states, which is valid at room temperature. Fluorescent excitation (decay) is indicated with green (red) arrows, non-radiative decay pathways are visualized with gray arrows (dashed/solid lines represent slow/fast decay channels). \textit{Right}: At intermediate and low temperatures, orbital hopping in the excited state (not spin conserving, black arrows) reduces spin contrast.  (b) Simulated photoluminescence dynamics during a readout laser pulse for an NV center in the $m_S = 0$ (dashed line) and $m_s = \pm 1$ (dotted line) spin state at three different temperatures (from left to right: 0 K, 150 K, 300 K). (c) Schematic visualization of the measurement protocol for nanoscale nuclear magnetic resonance of $^{13}$C nuclear spins (yellow) with single NV centers (blue). Polarization is transferred from the NV center to the nuclear spin (blue arrow), nuclear spin precession is initiated (yellow pulse). The nuclear spin precession (black oscillation trajectory) is tracked \textit{via} the NV center using a series of weak measurements (blue pulses). Each weak measurement block consists of optical initialization of the NV center (green pulse), an XY-8 dipolar pulse sequence (grey pulses) and optical NV readout. (d) The free induction decay (\textit{left}) and spectrum (\textit{right}) of the $^{13}$C nuclear spin is obtained as the combined result of all weak measurements. }
\label{fig:NVcenter}
\end{figure*}

The NV center is a spin defect in diamond with remarkably long coherence times even at room-temperature \cite{Du2024}. It offers all-optical spin state initialization and readout as well as coherent manipulation of spin states with resonant microwave fields \cite{Du2024}. \simos includes a sub-module with NV specific functionality including physical constants as well as pre-written routines for system construction and evolution. In the following, we showcase how to simulate  temperature-dependent  photophysics of the NV center as well as the coherent detection of nuclear spin-clusters by single NV centers with \simos. 

\subsection{Photophysics of the NV Center in Diamond}

The spin state of the negatively charged NV center is commonly optically prepared and read out using off-resonant green light illumination thus enabling single-spin detection spanning room temperature to cryogenic temperature \cite{Ernst2023}. To simulate the optical excitation and population dynamics of single NV centers, the electronic level structure must be taken into account. In the simplest case, valid at room temperature, the negatively charged NV center electronic structure may be described by three electronic levels: a triplet ground state, a triplet excited state and a metastable singlet state (Figure \ref{fig:NVcenter} (a)). The excited state has an optical lifetime of tens of nanoseconds and decays either radiatively by emission of a red photon, or non-radiatively \textit{via} an intersystem crossing (ISC) to the metastable singlet state. This ISC is strongly spin dependent, resulting in (i) higher fluorescence intensity for the $m_S = 0$ spin state versus the $m_S = 1$ spin state and (ii) build-up  $m_S = 0$ population after multiple excitation-decay cycles.   

The composite quantum system, which accounts for NV center spin and electronic structure, can be constructed in \simos as follows,
\begin{lstlisting}[language=Python]
# NV Center Electron Spin
S = {'val': 1, 'name':'S', 'type': 'NV-'} 
# Electronic States
GS = {'val': 0 , 'name':'GS'}
ES = {'val': 0 , 'name':'ES'}
SS = {'val': 0 , 'name':'SS'}
s = sos.System(([(GS, ES), S], SS))
\end{lstlisting}
Incoherent excitation and decay are defined by a rate dictionary that is passed on to the  \texttt{transition\_operator} routine. 
Alternatively, the NV sub-module provides the method \texttt{NVSystem} which initializes an NV system based on a series of user-defined options. A call
\begin{lstlisting}[language=Python]
s = sos.NV.NVSystem(nitrogen = False)
\end{lstlisting}
constructs the same system as defined above. Importantly, if \texttt{NVSystem} is being used, the collapse operators for optical transitions do not have to be defined manually, but can rather be extracted by
\begin{lstlisting}[language=Python]
c_ops_on, c_ops_off = s.get_transitionoperators(T=300, beta=0.2)
\end{lstlisting}
where \texttt{T} and \texttt{beta} define the system temperature and laser excitation power (saturation for \texttt{beta}=1) and \texttt{c\_ops\_on} and \texttt{c\_ops\_off} are the collapse operators in the presence and absence of laser illumination, respectively.

While this simple and fully incoherent model sufficiently describes NV center photophysics at elevated temperatures, it is invalid at cryogenic and intermediate temperatures below ca. \SI{100}{K}.  Here, interplay of spin and orbital dynamics in the excited state may result in fast spin relaxation and vanishing ODMR contrast \cite{Ernst2023}. In this regime, simulations must be performed with a quantum master equation that includes coherent spin dynamics under external magnetic and electric (strain) fields as well as incoherent excitation and decay dynamics accounting for the orbital character of the excited state and phonon-induced hopping \cite{Ernst2023}. We initialize the full system as
\begin{lstlisting}[language=Python]
s = sos.NV.NVSystem(nitrogen = False, orbital = True)
\end{lstlisting}
and define the system Hamiltonian and collapse operators for a specific temperature \texttt{T} and magnetic and electric (strain) fields \texttt{Bvec, Evec}  
\begin{lstlisting}[language=Python] 
HGS, HES = s.get_Hamiltonian(Bvec=Bvec,Evec=Evec)
H = HGS + HES
c_ops_on, c_ops_off = s.get_transitionoperators(T=300,  beta=0.2, Bvec=Bvec, Evec=Evec)
\end{lstlisting}
using the native \texttt{NVSystem} methods \texttt{get\_Hamitonian} and \texttt{get\_transitionoperators}.  Finally, we simulate the photoluminescence dynamics during a readout laser pulse for an NV center in the $m_s= 0$ and $m_S = 1$ state, respectively. 
\begin{lstlisting}[language=Python]
pl0 = []
pl1 = []
U  = sos.evol((HGS+HES), dt, c_ops = c_ops_on)
rho0 = (NV.GSid*NV.Sp[0]).copy()
rho1 = (NV.GSid*NV.Sp[1]).copy()
for t in tax:
    pl0.append(sos.expect(NV.ESid, rho_0))
    pl1.append(sos.expect(NV.ESid, rho_1))  
    rho0 = sos.applySuperoperator(U,rho0)
    rho1 = sos.applySuperoperator(U,rho1)
\end{lstlisting}
Figure \ref{fig:NVcenter} (b) shows the results for selected temperatures ranging from 0 to 300 K. Clearly, spin contrast is reduced at low temperatures, where phonon-induced orbital hopping is slow and induces spin-relaxation. It reaches a minimum for temperatures where the strain induced energy splitting of the excited state orbital branches matches the phonon induced transition rates \cite{Ernst2023}. At temperatures $>$ \SI{100}{K} the low-temperature model correctly approaches the simplified room-temperature model. %

\subsection{Weak Measurements with NV Centers in Diamond}

NV centers have been used to perform nanoscale NMR measurements of proximal nuclei both within the diamond lattice  and, using shallow defects, external nuclei in chemical targets \cite{Abobeih2019, Lochvinsky2016, Abendroth2022}.  The NV center is further capable of detecting the coherent precession of nearby nuclear spins using a measurement technique referred to as weak measurements \cite{Cujia2019, Herb2024}.  Here, the nuclear spin is polarized and rotated with a $\nicefrac{\pi}{2}$ pulse to initiate precession reminiscent of a free-induction decay in NMR. Repeated weak measurements on the NV center, consisting of optical spin initialization, XY-8 dynamical decoupling frequency encoding, and optical readout are used to track the nuclear spin precession (Figure \ref{fig:NVcenter} (c)). 

In the simplest case, assuming ideal optical initialization and readout, this measurement protocol can be simulated without considering the photophysics of the NV center. The quantum system is constructed from the NV center's electronic spin $S=\nicefrac{1}{2}$ which is coupled to a single $^{13}$C nuclear spin $S=\nicefrac{1}{2}$. Again, we use the NV sub-module method \texttt{NVSystem} to initialize the system. By default, the electronic levels of the NV center and a coupled nitrogen spin are taken into account. To exclude them, we set the \texttt{optics=False} and \texttt{nitrogen=False} options upon system construction. 
\begin{lstlisting}[language=Python]
C = {"name": "C", "val": 1/2}
NV = sos.NV.NVSystem(further_spins = [C], optics = False, nitrogen = False)
\end{lstlisting} 
In a next step, we prepare the initial state that represents the polarized and rotated $^{13}$C spin coupled to a polarized NV center and define the system Hamiltonian from the native system operators. We work in the rotating frame for the NV center where the hyperfine interaction is truncated to the secular (\texttt{apara} $a_{\parallel}$) and pseudo-secular (\texttt{aperp} $a_{\perp}$) contributions and utilize the $^{13}$C gyromagnetic ratio (\texttt{yC13}) of \simos. 
\begin{lstlisting}[language=Python]
# Initial State
rho0 = sos.state(s, "S[0],C[0.5]")
rho1 = sos.rot(s.Cx,np.pi/2,rho0)   
# System Hamiltonian
H0 = sos.yC13*B0*s.Cz + apara*s.Sz*s.Cz + aperp*s.Sx*s.Cx
\end{lstlisting}
Finally, we perform the sensing protocol, consisting of a series of \texttt{N} XY-8 blocks that are interspaced by free evolution periods. Again, we make use of the NV sub-module functionality and utilize the \texttt{XY8} and \texttt{meas\_NV} routines to apply the dipolar decoupling sequence  and perform NV center readout.
\begin{lstlisting}[language=Python]
N = 500
frf = sos.w2f(B0*sos.yC13)
twait = 0.125/frf
tau = 1/frf/2
store = []
rho = rho1.copy()
for i in range(N):
    rho = sos.rot(s.Sop_y_red,np.pi/2,rho)
    rho = sos.NV.XY8(H0,tau,s,rho,N=16)
    rho = sos.rot(s.Sop_x_red,np.pi/2,rho)
    meas,rho = sos.NV.meas_NV(rho,s)
    store.append(meas)
    rho = sos.evol(H0,twait,rho)
Tsample = 16*tau + twait
foffset = np.round(Tsample*frf)/Tsample
\end{lstlisting}
The resulting free-induction decay (contained in the variable \texttt{store} in our code example) and the spectrum that results after Fourier transformation are shown in Figure \ref{fig:NVcenter} (d).  Although not included here for simplicity, a variety of other effects may be incorporated in the simulation, e.g., spin relaxation, explicit simulation of nuclear polarization buildup, or details of the NV photophysics.

\section{Example 3: Spin-Correlated Radical Pairs (SCRPs)}
\label{sec:examplesSCRP}

\begin{figure*}
    \includegraphics[width=1.0\textwidth]{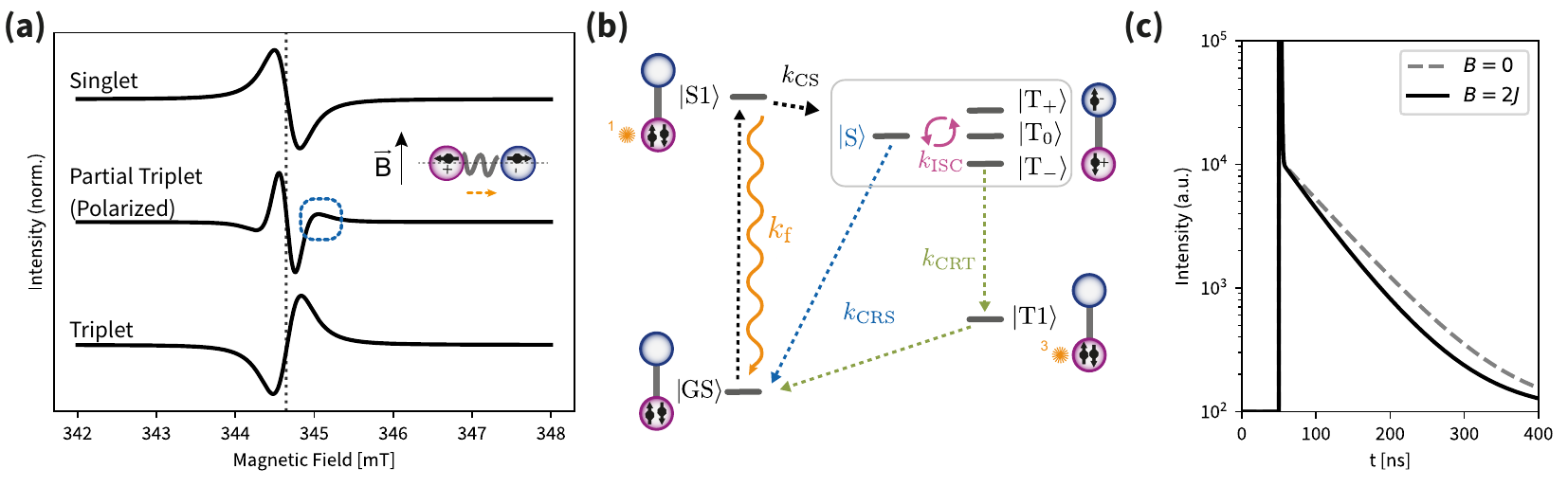}
    \caption{EPR spectroscopy (a) and ODMR spectroscopy (b, c) of radical pairs with \simos. (a) cw-EPR spectra of ensembles of photogenerated spin-correlated radical pairs (SCRPs) initially prepared in (\textit{top}) a singlet state, (\textit{middle}) a partially spin-polarized entangled state and (\textit{middle}) a triplet state. The SCRPs are oriented perpendicular with respect to an external field $\vec{B}$. The finite triplet character of the partially polarized state manifests itself in characteristic lineshapes for molecular ensembles oriented perpendicular to an external magnetic field ("wings", see dashed blue box). (b) Jab\l{}o\'nski diagram depicting photoinduced excitation and decay events relevant for generation and annihilation of entangled radical pairs. Excitation of a diamagnetic precursor ($\ket{\mathrm{GS}}\rightarrow\ket{\mathrm{S1}}$) is followed by spin-conserving charge separation and SCRP formation ($\ket{\mathrm{S1}}\rightarrow\ket{\mathrm{S}}$). Recombination of the SCRP occurs in a spin selective manner for singlet ($\ket{\mathrm{S1}}\rightarrow\ket{\mathrm{S}}$) and triplet ($\ket{\mathrm{T}}\rightarrow\ket{\mathrm{T1}}\rightarrow\ket{\mathrm{GS}}$) spin states. (c) Time-resolved fluorescence of the radical pair during and after a short laser pulse at two different magnetic fields $B$ on and off the $2J$ resonance. At the $2J$ resonance enhanced singlet--triplet interconversion lowers fluorescence lifetime.}
    \label{fig:SCRP}
\end{figure*}

Spin-correlated radical pairs (SCRPs) are electron spin pairs generated  by photoexcitation and subsequent electron transfer. As electron transfer is spin conserving, SCRPs are initiated in well-defined, entangled spin states. Spin chemistry of SCRPs plays important roles in biology, e.g., in solar energy harvesting by photosynthetic reaction centers \cite{THURNAUR1980557}, and has been implicated in the leading hypothesis for avian magnetoreception \textit{via} the radical pair mechanism in blue-light sensitive cryptochromes \cite{Rodgers2009,Xu2021}. Synthetic molecules mimicking these biologically relevant species have garnered interest as chemical approaches to quantum sensing and quantum information science. \cite{Mani2022} Further they are being explored to probe to the potential role of chirality-induced spin selectivity (CISS) on the initial spin state and charge recombination of SCRPs. \cite{Fay2021,Voelker2023, chiesa2023AdvMater}
Characterization techniques of SCRPs include EPR as well as optically or chemically detected magnetic resonance. A specific submodule is available in \simos that provides high-level functions to facilitate simulations of common SCRP characterization techniques. In the following, we demonstrate simulations of continuous wave EPR and ODMR of magnetic field effects for two selected examples of recently published original research.

\subsection{Continuous-Wave EPR of SCRPs}

Continuous-wave (cw) EPR spectroscopy has been used to detect a partial triplet character in the initial state of photogenerated SCRPs formed in synthetic donor--bridge--acceptor molecules with chiral bridge units \cite{Eckvahl2023}. The triplet character in the initial spin state of the SCRPs, attributed to CISS-induced spin polarization, is measured by cw EPR of SCRP samples oriented perpendicular to the external magnetic field. To qualitatively reproduce the recent findings of Eckvahl \textit{et al.}\cite{Eckvahl2023}, we construct a SCRP system consisting of two electron and two nuclear spins,
\begin{lstlisting}[language=Python]
# Electron spin 1 
A = {'val': 1/2, 'name':'A'}  
# Electron spin 2
B = {'val': 1/2 , 'name':'B'} 
# Coupled nuclear spins 
HA = {'val': 1/2 , 'name':'HA'}
HB = {'val': 1/2 , 'name':'HB'}
s = sos.System([A, B, HA, HB])
\end{lstlisting}
We define all interactions for both, parallel and perpendicular oriented radical pairs, including anisotropic Zeeman-interactions of the electron spins, a distribution of dipolar couplings between electron spins, and hyperfine interactions between electron and nuclear spins. To obtain the initial states of radical pairs, we utilize the \texttt{state} function of the SCRP submodule. Here, the coherent initial state of the SCRP is parameterized with a minimal set of parameters, $\alpha, \beta$ \cite{Voelker2023} 
\begin{equation}
\ket{\psi_{\parallel}} = \cos{\alpha}\ket{S} + e^{i \beta}\sin{\alpha}\ket{T_0}.
\label{eq:initial_state}
\end{equation}
The relative orientation of the SCRP with respect to the designated field axis may be specified  using the \texttt{state} method. To compare the spectra of SCRPs with different spin states oriented perpendicular to the external field, we define initial states for a series of  $\alpha = 0^{\circ}, \sim  20^{\circ},  90^{\circ}$ and otherwise identical parameters (i.e. $\beta = 0^{\circ}$ and $\theta = 90^{\circ}$).
 \begin{lstlisting}[language=Python]
rho0 = sos.SCRP.initial_state(s, "A", "B", alpha, 0, 1, theta = theta, phi = 0)
 \end{lstlisting}
Finally we simulate the field sweep, producing cw-EPR spectra for both orientations using the high-level \texttt{cwEPR\_fieldsweep} method of the SCRP submodule. 
 \begin{lstlisting}[language=Python]
axis, spectrum = sos.SCRP.cwEPR_fieldweep(s, ["A", "B"],  rho0, Hfield , Hrest, sos.f2w(9.67e9), broadening = 0.5e-3)
 \end{lstlisting}
 In contrast to pulsed EPR techniques, simulation of cw-EPR spectra does not require explicit propagation of an initial state. Rather, \texttt{cwEPR\_fieldsweep} determines resonant fields \textit{via} matrix diagonalization in a higher-dimensional, Liouville-type space and obtains intensities as products of population differences and transition probabilities. Input arguments for \texttt{cwEPR\_fieldsweep} must include 
\begin{enumerate}
\item a list of all spins that should be considered in the simulation (i.e., \texttt{["A", "B"]} 
\item the initial state of the SCRP   \texttt{rho0}
\item the field-dependent part of the Hamiltonian \texttt{Hfield} 
\item the field-independent part of the Hamiltonian \texttt{Hrest}
\item the microwave frequency (here, \SI{9.67}{GHz}, in the code above we utilize the \texttt{f2w} method of \simos \texttt{f2w} to convert to angular frequency units)
\item the extent of spectral broadening.
\end{enumerate}
 Figure \ref{fig:SCRP} (a) visualizes the calculated cw EPR spectra of the different spin states, revealing distinct spectral line-shapes indicative of partial spin polarization due to CISS.

\subsection{Magnetic Field Effects}
Together, cw and pulsed EPR methods enable the direct observation of spin--field and spin--spin interactions of SCRPs with high frequency resolution. However, experiments suffer from an inherently low sensitivity, requiring large sample quantities, high magnetic fields of several hundred mT, or, in some cases, cryogenic conditions. ODMR methods, in contrast, offer higher sensitivity and enable characterization of SCRPs at room temperature. Here, an optical signal, e.g., transient absorption or fluorescence intensity, is detected as a function of the external magnetic field strength or the frequency of an applied microwave field. The former is referred to as magnetic-affected reaction yield (MARY) while the latter is commonly termed reaction yield detected magnetic resonance (RYDMR) \cite{Mani2022}. Unless specific conditions are met, simulation of MARY or RYDMR spectra requires incorporating the incoherent excitation and decay dynamics of the SCRP in addition to coherent spin dynamics in the charge-separated state. As visualized in Figure \ref{fig:SCRP} (b), these include photo-excitation of the diamagnetic precursor, charge separation, and spin-selective recombination (or reaction). 
For our illustration, we demonstrate how to simulate the magnetic-field affected fluorescence lifetime of a SCRP system following recent work by Buck \textit{et al.} \cite{Buck2020}. We start by initializing the system, including the electron spin pair, a  nuclear spin coupled to one of the electrons and a series of electronic levels (i.e., electronic ground and excited singlet states, the charge-separated state and an excited triplet state). We also introduce the singlet and triplet basis states of the electron spin pair.
 \begin{lstlisting}[language=Python]
# Electron spin 1
A = {'val': 1/2, 'name':'A'} 
# Electron spin 2 
B = {'val': 1/2 , 'name':'B'} 
# Nuclear spin 
H = {'val': 1/2 , 'name':'H'} 
# Electronic ground state
GS =  {'val': 0 , 'name':'GS'} 
# Excited state 
S1 =  {'val': 0 , 'name':'S1'} 
# Intermediate state for triplet
# recombination 
T1 =  {'val': 0 , 'name':'T1'} 
# Construct the system. 
rp = sos.System((GS, S1, [A,B, H], T1), method = "qutip")
# Singlet/Triplet Ghostspins 
rp.add_ghostspin("C", ["A", "B"])
 \end{lstlisting}
In a second step we define the coherent system Hamiltonian as well as the collapse operator for photo-excitation and successive decay paths. A code snippet below demonstrates how to set up the rate dictionary for optical excitation and decay channels. 
 \begin{lstlisting}[language=Python]
rates = {}
rates_laser = {}
rates_laser["GS->S1"]  = beta*kfl
rates["S1->GS"] = kfl
rates["S1->C_1[0]"] = kcs
rates["S1<-C_1[0]"] =  kbcr
rates["C_1[0] -> GS"] = kcrs
rates["C_3[-1]-> T1"] = kcrt
rates["C_3[0] -> T1"] = kcrt
rates["C_3[1] -> T1"] = kcrt
rates["T1 -> GS"] = kcrtt
\end{lstlisting}
Finally we evaluate the fluorescence response during and after a short laser pulse at two distinct magnetic fields, on and off the $2J$ resonance where magnetic field strength equals twice the scalar $J$ coupling of the SCRP. 
 \begin{lstlisting}[language=Python]
Bs = [0.5e-3, 100e-3] # Tesla
rho0 = rp.GSid.unit() 
pl = rp.S1id 
for ind_B, B in enumerate(Bs):
    H = Hhfi + HJ +  HZ(B) 
    Ubright = sos.evol(H, dt, c_ops = c_ops_on) 
    Udark = sos.evol(H, dt, c_ops = c_ops_off) 
    rho = rho0.copy()
    for i in range(pts): 
        meas = sos.expect(pl ,rho)
        luminescence[ind_B, i] = meas 
        # Laser Pulse
        if  i in range(start, stop):
            rho = sos.applySuperoperator(Ubright,rho) 
        # Before/After Laser Pulse
        else:
            rho = sos.applySuperoperator(Udark,rho)
\end{lstlisting}

As visualized in Figure \ref{fig:SCRP}(d), reduced fluorescence lifetimes are observed at the $2J$ resonance due to facilitated singlet--triplet interconversion.

\section{Example 4: NMR Powder averages and MAS with the Fokker-Planck Module}
\label{sec:fokkerexample}

\begin{figure}
    \centering
    \includegraphics[width=\linewidth]{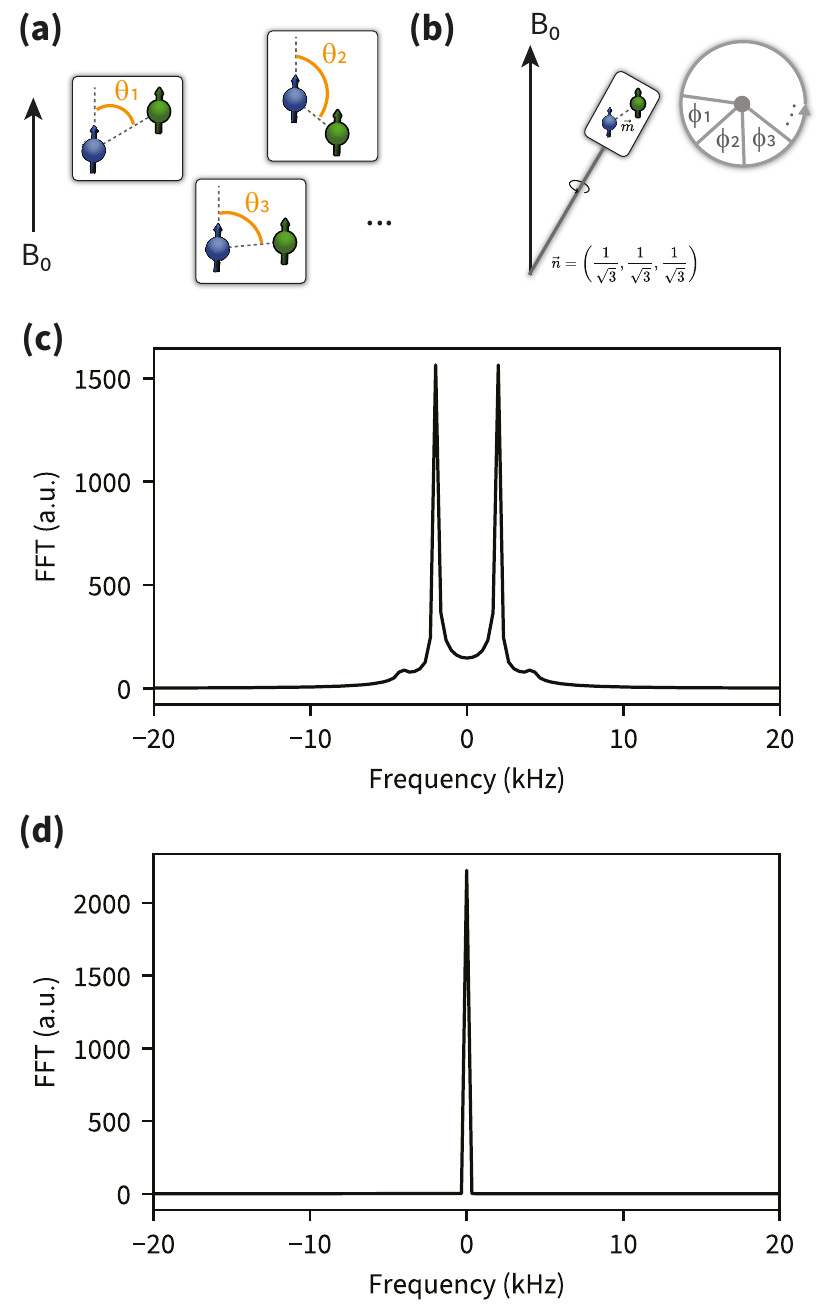}
    \caption{Simulation of ssNMR using \simos. (a) shows the situation of powder average of a spin pair: Different orientations $\theta_i$ lead to a varying coupling between spin pairs of the ensemble. (b) shows the situation for magic angle spinning: The sample is rotaed around an axis $\vec{n}$ that is tilted by 54.7\textdegree~ relative to the $B_0$ field. This leads to rotor phase angles $\phi_i$ that modulate the coupling of the spin pair. (c) Without MAS, a powder average leads to the famous Pake pattern. (d) The use of MAS decouples anisotropic interactions and therefore narrows the spectrum, here by suppressing the dipolar coupling of the spin pair.}
    \label{fig:fokker_demo}
\end{figure}

We illustrate the syntax of the Fokker--Planck module by simulating the powder-average of a dipolar-coupled, heteronuclear spin pair ($I$, $J$) ensemble under static and magic angle spinning (MAS) conditions.
Individual spin pairs within the powder have different orientations $\theta$ relative to the spin-aligning magnetic field $B_0$ and, consequently, different dipolar couplings. The NMR signal is an average over all orientations. In the static case, the spectrum a characteristic shape commonly referred to as a \textit{Pake pattern}. Under MAS at speeds exceeding the dipolar coupling strength, the Pake pattern  vanishes and the NMR spectrum reduces to a single peak, reflecting the time-averaged effective Hamiltonian.\cite{Bockmann2015}

To simulate a powder average, we parameterize the stochastic part of the Hamiltonian, namely the dipolar coupling interaction, and define basis functions of the stochastic space to use with the \texttt{stochastic\_evol} routine. We implement a function that returns the dipolar coupling Hamiltonian for a given orientation $\theta$,
\begin{lstlisting}[language=Python]
def H_fun(theta):
  r = np.array([3e-10,theta,0]
  H = sos.dipolar_coupling(s, "I", "J", sos.yH1, sos.yF19, r, approx = "secular")
  return Hd
\end{lstlisting}
and define a \texttt{StochasticLiouvilleParameters} object that holds the discrete values and weights of the stochastically modulated parameter $\theta$. 
\begin{lstlisting}[language=python]
para = sos.StochasticLiouvilleParameters()
para["theta"].values = np.linspace(0,np.pi,100)
para["theta"].weights = np.sin(para["theta"].values)
\end{lstlisting}

In a next step, we prepare the initial state and transform it into Fokker--Planck space.  We use the \texttt{pol\_spin} routine of \simos which initializes  $S = \nicefrac{1}{2}$ spins in a state with given polarization, prepares a coherent state with a $\nicefrac{\pi}{2}$ pulse, and transforms the resulting density matrix into a vectorized Fokker--Planck representation with the \texttt{dm2f} routine. Note that detection is performed on spin $I$. Spin $J$ is assumed to be in a thermal state. 
\begin{lstlisting}
I0 = sos.pol_spin(1)
J0 = sos.pol_spin(0)
rho0 = sos.tensor([I0, J0])
rho = sos.rot((s.Ix), np.pi/2, rho0)
rhovec = sos.dm2fp(rho, para.dof)
\end{lstlisting}
Next we execute the simulation. For more efficient computation, we can cache the propagator $U$,
\begin{lstlisting}
U = sos.stochastic_evol(H_fun, para, dt, method='qutip',space='liouville')
\end{lstlisting}
The time trace is obtained by applying $U$ repeatedly to \texttt{rhovec} and evaluating the expectation value of \texttt{s.Sx+1j*s.Sy}. Fourier transformation results in the expected Pake pattern, shown in Figure \ref{fig:fokker_demo} (c)).

In MAS, the sample is rotated around an axis which stands at the so-called magic angle of $\alpha=54.7\text{\textdegree}$ relative to the magnetic field axis in order to average out dipolar broadening. To simulate MAS we modify the dipolar Hamiltonian function to incorporate the rotor phase $\phi$.
\begin{lstlisting}
from scipy.spatial.transform import Rotation
def H_fun_MAS(phi):
  m = sos.spher2cart(np.array([3e-10, 0, 0]))
  rotvec = phi*sos.mas_axis
  R = Rotation.from_rotvec(rotvec)
  Hdip = sos.dipolar_coupling(s, "I", "J", sos.yH1, sos.yF19, R.apply(m), mode = "cart", approx = "secular")
  return Hdip 
\end{lstlisting}
We define a \texttt{StochasticLiouvilleParameter} object to discretize the rotor phase and define values and weights. Further, we introduce the stochastic dynamics and specify linear (i.e., a first order) dynamics by choosing a MAS frequency (in the following example set to \SI{100}{kHz}). 
\begin{lstlisting}
para = sos.StochasticLiouvilleParameters()
para["phi"].values = np.linspace(0,np.pi,100)
para["phi"].dynamics = {1:sos.f2w(100e3)}
para["phi"].weights = None
\end{lstlisting}
Next, we use the \texttt{simos.stochastic\_evol} method to simulate the time evolution of the quantum system. Note that we sample the time trace in a rotor-synchronized fashion. As shown in Figure \ref{fig:fokker_demo} (d), the dipolar coupling vanishes and the spectrum reduces to a single peak.

\section{Installation and Virtual Lab}
\simos is available on Github, or on the PyPI package repository. Installation instructions can be found in the documentation available at Read The Docs \cite{doc}. On Github, extensive example Notebooks are available including all demonstrations discussed in this paper in addition to many more examples. Furthermore, we provide a browser-based environment that is ready to use at \url{https://simos.kherb.io}. Thanks to the Pyodide project \cite{pyodide}, \simos with all its dependencies can be compiled to WebAssembly. Thus, \simos may be run fully in the browser without installing, while keeping all code execution local for performance and privacy reasons. While installation is always recommended, the web-based implementation enables a quick try of many of \simos's functionality including all example notebooks.

\section{Conclusion}
\simos is a toolkit for performing magnetic resonance simulations in Python. While developed with  simulation of optically-addressable spins in mind, \simos is also a fully fledged simulation suite for NMR and EPR experiments in Python. The authors would like to note that no Python-based library exists to date for this purpose. Due to its backend-agnostic approach, simulations can and will be easily ported to more frameworks such as PyTorch or CuPy in the future. The clear focus of \simos on a clean and mathematical syntax while adhering to a Pythonic way of programming distinguishes it from other packages.  

\section*{Acknowledgment}
The authors thank Ilia Solov'yov and Luca Gerhards for fruitful discussions. L.A.V. and J.M.A. acknowledge funding from the Swiss National Science Foundation (SNSF) Ambizione Project Grant No. PZ00P2-201590. K.H. and C.L.D. acknowledge funding from the SBFI, Project "QMetMuFuSP" under Grant. No. UeM019-8, 215927 and from the SNSF, under Grants. No. 200021-219386 and CRSII-222812.
\label{sec:final}

\section*{Author contributions}
L.A.V. and K.H designed and implemented all program routines based on a prototype implemented by K.H. \cite{herb_master}. The manuscript was co-written by L.A.V. and K.H. with assistance of J.M.A. and C.L.D.

\bibliographystyle{elsarticle-num}
\bibliography{references}

\end{document}